\documentclass[letters,usenatbib]{mnras}

\usepackage[T1]{fontenc}
\usepackage{ae,aecompl}

\usepackage{graphicx}	
\usepackage{amsmath}	
\usepackage{amssymb}	

\title[Quenching Evolution]{Evidence for strong evolution in galaxy environmental quenching efficiency between z = 1.6 and z = 0.9}

\author[Nantais et al.]{
Julie B. Nantais,$^{1}$\thanks{E-mail: julie.nantais@unab.cl}
Adam Muzzin,$^{2}$
Remco F. J. van der Burg,$^{3}$
Gillian Wilson,$^{4}$
\newauthor
Chris Lidman,$^{5}$
Ryan Foltz,$^{4}$
Andrew DeGroot,$^{4}$
Allison Noble,$^{6}$
Michael C. Cooper,$^{7}$
\newauthor
and Ricardo Demarco$^{8}$\\
$^{1}$Departamento de Ciencias F\'isicas, Universidad Andres Bello, Fernandez Concha 700, Las Condes 7591538, Santiago, \\
Regi\'on Metropolitana, Chile\\
$^{2}$Institute of Astronomy, University of Cambridge, Madingley Road, Cambridge CB3 0HA, United Kingdom\\
$^{3}$Laboratoire AIM-Paris-Saclay, CEA/DSM-CNRS-Universit\'e Paris Diderot, Irfu/Service d'Astrophysique, CEA Saclay, Orme des Merisiers,\\ 
91191 Gif-sur-Yvette, France\\
$^{4}$Department of Physics and Astronomy, University of California-Riverside, 900 University Avenue, Riverside, CA 92521, USA\\
$^{5}$Australian Astronomical Observatory, PO Box 2915, North Ryde NSW 1670, Australia \\
$^{6}$Department of Astronomy \& Astrophysics, University of Toronto, Toronto, Ontario M5S 3H4, Canada\\
$^{7}$Department of Physics and Astronomy, University of California, Irvine, 4129 Frederick Reines Hall, Irvine, CA 92697, USA\\
$^{8}$Departamento de Astronom\'ia, Universidad de Concepci\'on, Casilla 160-C, Concepci\'on, Regi\'on del Biob\'io, Chile}

\date{Accepted XXX. Received YYY; in original form ZZZ}

\pubyear{2016}

\begin{document}
\label{firstpage}
\pagerange{\pageref{firstpage}--\pageref{lastpage}}
\maketitle

\begin{abstract}
We analyse the evolution of environmental quenching efficiency, the fraction of quenched cluster galaxies that would be star-forming if they were in the field, as a function of redshift in 14 spectroscopically confirmed galaxy clusters with 0.87 $<$ $z$ $<$ 1.63 from the $Spitzer$ Adaptation of the Red-Sequence Cluster Survey (SpARCS).  The clusters are the richest in the survey at each redshift.  Passive fractions rise from $42_{-13}^{+10}$\% at $z$ $\sim$ 1.6 to $80_{-9}^{+12}$\% at $z$ $\sim$ 1.3 and $88_{-3}^{+4}$\% at $z$ $<$ 1.1, outpacing the change in passive fraction in the field.  Environmental quenching efficiency rises dramatically from $16_{-19}^{+15}$ at $z$ $\sim$ 1.6 to $62_{-15}^{+21}\%$ at $z$ $\sim$ 1.3 and $73_{-7}^{+8}$\% at $z$ $\lesssim$ 1.1.  This work is the first to show direct observational evidence for a rapid increase in the strength of environmental quenching in galaxy clusters at $z$ $\sim$ 1.5, where simulations show cluster-mass halos undergo non-linear collapse and virialisation.
\end{abstract}




\section{Introduction}

The history of how galaxy clusters came to be dominated by passive, early-type galaxies (Dressler 1980) is a long-standing problem. Over cosmic time, protoclusters, dominated by high-mass star-forming galaxies at $z$ $>$ 2 (Overzier et al.~2008; Galametz et al.~2010; Hatch et al.~2011a, 2011b; Shimakawa et al.~2014; Umehata et al.~2015), must evolve into mature galaxy clusters with well-established red sequences at $z$ $<$ 1 (Muzzin et al.~2012; Foltz et al.~2015; Balogh et al.~2016).  This observed evolution in the cluster population at 1 $<$ $z$ $<$ 3 suggests a corresponding rapid increase in the environmental quenching efficiency, which is defined to be the fraction of passive group or cluster galaxies that are would be star-forming if they were in the field.  Cluster galaxies at $z$ $\sim$ 1 are most likely to be quenched if they have spent significant time in the cluster (Stanford, Eisenhardt, \& Dickinson 1998; Blakeslee et al.~2006; Demarco et al.~2010; Muzzin et al.~2012, 2014; Nantais et al.~2013a,b; Noble et al.~2013, 2016), indicating that the cluster environment is important for transforming galaxies from star-forming to passive.  

However, the redshift gap between clusters and protoclusters is only beginning to be bridged, and the transition between the two cannot necessarily be attributed to a single obvious factor such as redshift or cluster mass.  For instance, the $z$ $\sim$ 2.5 protocluster in Wang et al.~(2016) appears to be as massive as many rich $z$ $\sim$ 1.5 galaxy clusters, but is still rich in high-mass dusty star-forming galaxies like other $z$ $>$ 2 protoclusters.  Also, there is considerable variation even among galaxy clusters at 1.3 $<$ $z$ $<$ 2, with many claiming notable enhancement in quenching (Kodama et al.~2007, Bauer et al.~2011; Quadri et al.~2012; Gobat et al.~2013; Strazzullo et al.~2013; Andreon et al.~2014; Newman et al.~2014; Balogh et al.~2016; Cooke et al.~2016) while others note still-substantial star formation among massive galaxies (Brodwin et al.~2013; Fassbender et al.~2014; Bayliss et al.~2014; Webb et al.~2015a,b; Bonaventura et al.~2016, submitted).  Studies of groups and clusters in large, general surveys, such as Gerke et al.~(2007), Cooper et al.~(2007), Kawinwanichakij et al.~(2016), and Darvish et al.~(2016), show greater decline in group passive fractions than in the field at $z$ $\sim$ 1.5 compared to $z$ $\sim$ 1, thereby showing signs of evolution in quenching efficiency.  However, these studies barely probe the very densest environments of rich galaxy clusters.

In this Letter, we use a sample of 14 spectroscopically confirmed galaxy clusters from the {\it{Spitzer}} Adaptation of the Red-Sequence Cluster Survey (SpARCS; Wilson et al.~2009; Muzzin et al.~2009) at 0.87 $<$ $z$ $<$ 1.63 to study evolution in environmental quenching efficiency as rich galaxy clusters assemble.  This work builds on Nantais et al.~(2016, hereafter N+16a) by using homogeneous processing at all redshifts and splitting clusters into similar co-moving volume bins.   In Section 2 we briefly describe the data and analysis methods used in our study.  In Section 3 we describe the results, in Section 4 we discuss the implications of the results, and in Section 5 we provide a brief summary.  We assume a $\Lambda$CDM cosmology with H$_0$ = 70 km s$^{-1}$ Mpc$^{-1}$, $\Omega_M$ = 0.3, and $\Omega_{\Lambda}$ = 0.7.

\section{Data and analysis}

Our sample contains 14 spectroscopically-confirmed near-infrared-selected clusters at 0.869 $<$ $z$ $<$ 1.633. The full redshift range of our clusters spans about 2.5 Gyr in cosmic time.  SpARCS covers an area on the sky of 41.9 deg$^2$ (Wilson et al.~2009), corresponding to a comoving volume of about 310 million Mpc$^3$ (see Table 1) between $z$ = 0.85 and $z$ = 1.65.  The multi-band photometry and spectroscopic members for the 10 clusters at $z$ $<$ 1.35 (GCLASS sample; Muzzin et al.~2012) are described in detail in van der Burg et al.~(2013, hereafter vdB+13). The GCLASS clusters range from $10^{14}$ to 2.6 $\times$ $10^{15}$ M$_{\sun}$ (vdB+13, see also Biviano et al.~2016). 

More details regarding the high-redshift (z $>$ 1.35) objects are described in N+16a.  In particular, the details of the broad-band photometry may be found in N+16a. All four N+16a clusters were targeted for spectroscopic confirmation as being the most significant overdensities as identified using photometric redshifts (Muzzin et al.~2013a).  Rough preliminary estimates of the masses for the two highest-redshift clusters in N+16a are given in Lidman et al.~(2012).  The two preliminary mass estimates in Lidman et al.~(2012) for N+16a clusters are consistent in their projected mass growth with vdB+13 clusters.  Richness-based mass estimates for three of the N+16 clusters in Delahaye et al.~(in preparation) range from $10^{14.3}$ to $10^{15}$ M$_{\sun}$.  No further radial velocity-based mass estimates are performed since these clusters are almost certainly not fully virialised, and because the spectroscopic instruments used to confirm the clusters bias our spectroscopic member sample toward unobscured star-forming galaxies.

The UltraVISTA/COSMOS field galaxy comparison data come from the 30-passband photometric catalogues of Muzzin et al.~(2013b), and are processed in a similar way to the cluster catalogues. We used EAZY (Brammer et al.~2008) to obtain photometric redshifts and the rest-frame $U-V$ and $V-J$ colours needed to estimate passive and star-forming fractions (Wuyts et al.~2007).  We use the Whitaker et al.~(2011) $UVJ$ criterion, as in vdB+13. We used FAST (Kriek et al.~2009) to derive stellar masses, using the same sets of models and parameters for GCLASS and high-redshift clusters.  Photometric cluster membership and $UVJ$ passive and star-forming status are defined as in vdB+13 and N+16a.  The photometric cluster membership criterion, $(z_{phot}-z_{cluster})/(1+z_{cluster})$ $\leq$ 0.05,  is similar to the normalised median absolute deviation (NMAD) scatter in the photometric redshifts: $\sigma$$\sim$0.03 for GCLASS and $\sigma$$\sim$0.04 for the N+16a clusters.  Whenever possible, we use spectroscopic redshifts to determine membership as in N+16a and vdB+13.

Cluster redshifts are binned into three redshift ranges for analysis: 0.85 $<$ $z$ $<$ 1.1, 1.1 $<$ $z$ $<$ 1.4 (including the N+16a cluster at $z$ = 1.37), and 1.4 $<$ $z$ $<$ 1.65 (therefore excluding the N+16a cluster at $z$ = 1.37).  Our different binning of the clusters is one of the major differences from N+16a that may lead to different results, the other being the use of homogeneous analysis methods at all redshifts for SpARCS clusters.  In each of these bins, at least one cluster is complete down to a lower stellar mass limit of 10$^{10.3}$ M$_{\sun}$.  This redshift binning was chosen primarily due to the similarity of the survey volume covered in each redshift range: 83, 121, and 113 million Mpc$^3$, respectively. Our binning also helps reduce the statistical uncertainties due to cluster-to-cluster variations at $z$ $<$ 1.5.

The growth of clusters between redshift bins in our sample is about two-thirds that predicted by theory, according to Lidman et al.~(2012).  This implies that the higher-redshift clusters may in fact be progenitors of somewhat more massive systems than some of the lower-redshift clusters, making any differences in environmental quenching with redshift less expected on theoretical grounds than would otherwise be the case.

Each cluster has a different 80\% stellar mass completeness limit from vdB+13 or N+16a, based on the lowest-mass passive galaxy visible at each $Ks$ limit to preserve accurate passive fractions.  Some are not complete to 10$^{10.3}$ M$_{\sun}$, the lowest UltraVISTA passive stellar mass limit at $z$ = 1.6.  Therefore, the total galaxy counts, as well as the individual passive and star-forming counts, in each cluster redshift bin must be corrected for this incompleteness. We count galaxies in each cluster down to that cluster's respective completeness limit or, if the photometry is deep enough, to 10$^{10.3}$ M$_{\sun}$.  Low-mass galaxy counts, those between 10$^{10.3}$ M$_{\sun}$ and 10$^{10.53}$ M$_{\sun}$ (the shallowest stellar mass limit reached by all clusters), are then adjusted by the inverse of the fraction of clusters complete at each low stellar mass, as in vdB+13 and N+16a.  Since the error sources are the same as in the stellar mass functions, we do not explicitly need to include stellar mass function uncertainties.  

After the adjustments for completeness in each redshift range, the total, $UVJ$ passive, and $UVJ$ star-forming fractions are summed in all the stellar mass bins to give the total counts, passive counts, and star-forming counts for a given cluster redshift range.  Since there is no evidence for evolution of the global stellar mass function in clusters between $z$ $\sim$ 1 and $z$ $\sim$ 1.5 (Andreon 2013; vdB+13; N+16a), nor is there evidence for significant global stellar mass function differences between clusters and the field at any redshift (Andreon 2013; vdB+13; N+16a; Vulcani et al. 2012, 2013, 2014), we use the same lower stellar mass limit at all redshifts and all environments in this work.

Since the highest-redshift (N+16a) clusters have almost entirely star-forming spectroscopic members due to instrument constraints, making adjustment for field contamination based on spectroscopy unfeasible, the adjustment of total cluster galaxy counts and passive fractions for field contamination is performed via subtraction of the expected contamination levels estimated from UltraVISTA/COSMOS (Muzzin et al.~2013b).  For each redshift bin, we subtract the number of field galaxies in the cluster photometric redshift range in the equivalent volume of the 3-6 SpARCS clusters in that redshift bin.  We estimate this volume by multiplying the area in Mpc$^2$ of each SpARCS cluster by the distance in Mpc between two non-gravitationally-bound galaxies on opposite ends of the cluster's photometric redshift selection range, which depends on the cluster's spectroscopic redshift.  The correction is applied separately to passive and star-forming galaxies so as to be able to correct the passive fraction and quenching efficiency for field interlopers.

Field galaxy passive fractions are also estimated from UltraVISTA as in N+16a, selecting from the full photometric redshift range corresponding to each redshift bin.  Total counts and passive counts are determined in these redshift ranges, and Poisson uncertainties are determined from these counts.

The environmental quenching efficiency, or conversion fraction, is the fraction of galaxies that would normally be star-forming in the field that are quenched in a group or cluster.  It is defined in the literature (van den Bosch et al.~2008, Peng et al.~2012, Phillips et al.~2014) as: $f_{conv} = (f_{p,cluster}-f_{p,field})/(f_{sf,field})$.  Total uncertainties in both cluster passive fractions and quenching efficiencies are derived from the 68\% confidence interval around the median of the probability density function of 100,000 Monte Carlo simulations, pairing 100 Monte Carlo simulations of the photometry randomly varied with Gaussian distributions around the photometric error bars with 1000 random UltraVISTA field samples.

For the four N+16a clusters, we consider only galaxies within 1 Mpc of the cluster centers, for consistency with vdB+13 and N+16a, and because reliable virial radius estimates are not available for these clusters. For the GCLASS clusters, we include galaxies out to the individual virial radius of the cluster as recalculated in Biviano et al.~(2016).

\begin{figure}
	\includegraphics[width=\columnwidth]{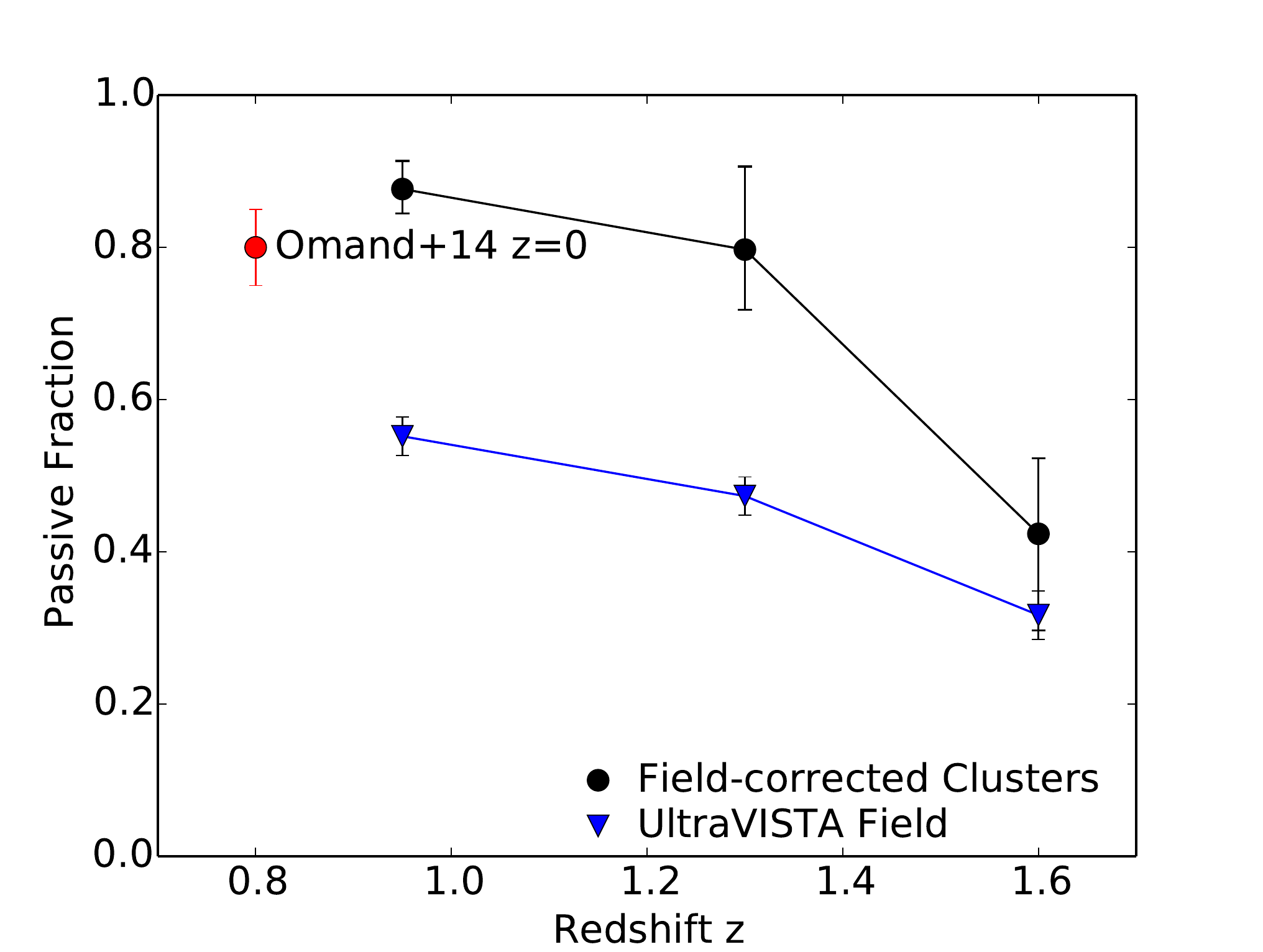}
    \caption{Passive fraction as a function of redshift for UltraVISTA/COSMOS field galaxies and cluster members with stellar masses of log (M*/M$_{\sun}$) $\geq$ 10.3 in three cluster redshift bins (1.4 $<$ $z$ $<$ 1.65, 1.1 $<$ $z$ $<$ 1.4, and 0.86 $<$ $z$ $<$ 1.1).  Black symbols represent cluster galaxies, and blue triangles are UltraVISTA/COSMOS field galaxies.  The red point shows the passive fraction at $z$ $=$ 0 from Omand et al.~(2014).}
    \label{qes}
\end{figure}

\begin{figure}
	\includegraphics[width=\columnwidth]{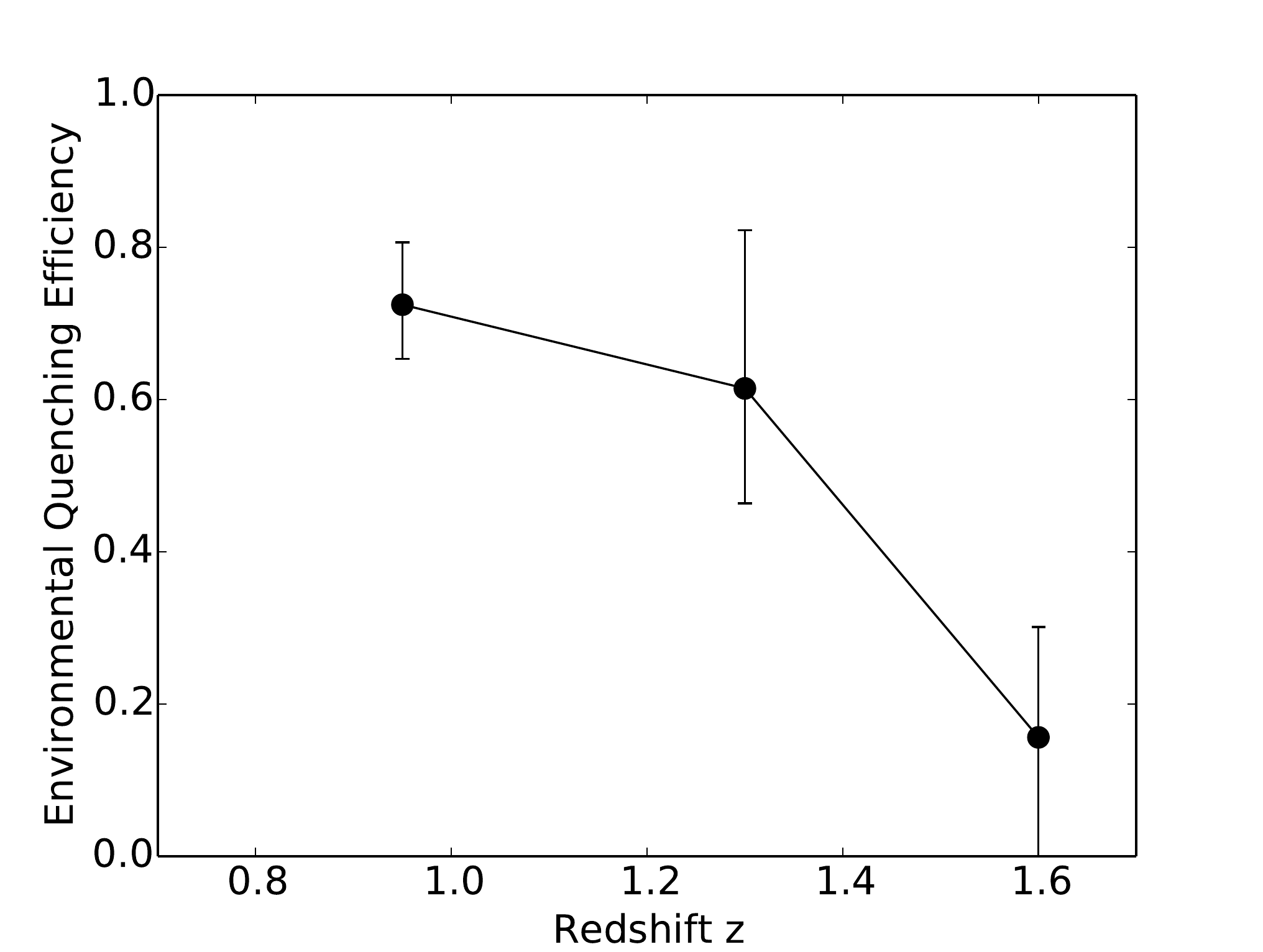}
    \caption{Environmental quenching efficiency of cluster galaxies with log (M$_*$/M$_{\sun}$) $\geq$ 10.3 as a function of cluster redshift for three cluster redshift bins (1.4 $<$ $z$ $<$ 1.65, 1.1 $<$ $z$ $<$ 1.4, and 0.86 $<$ $z$ $<$ 1.1).}
    \label{passfrac}
\end{figure}

\section{Results}

Figure 1 shows the passive fractions in our clusters as a function of redshift, along with the UltraVISTA field controls in blue.  At all redshifts, the passive fraction in clusters is higher than in the field.  The passive fraction nearly doubles from $z$ $\sim$ 1.6 to $z$ $\sim$ 1.3, jumping from 42\% to over 80\%, but changes little between $z$ $\sim$ 1.3 and $z$ $\lesssim$ 1.  The lower two redshift ranges have passive fractions consistent with the $z$ = 0 value from Omand et al.~(2014).  In the field, on the other hand, the passive fraction is only $\sim$ 50\% higher at $z$ $\sim$ 1.3 than at $z$ $\sim$ 1.6, and does not nearly double until $z$ $\lesssim$ 1.  Due to the large overdensities of our clusters, our results are still qualitatively robust, even without making a field galaxy background correction.

In Figure 2, we compare the environmental quenching efficiencies in the clusters as a function of redshift for all galaxies above 10$^{10.3}$ M$_{\sun}$.  For $z$ $\sim$ 1.6, the environmental quenching efficiency is consistent with zero ($16_{-19}^{+15}$\%), while it rises dramatically to $62_{-15}^{+21}$\% at $z$ $\sim$ 1.3 and $73_{-7}^{+8}$\% at $z$ $\lesssim$ 1.1. Our quenching efficiencies at $z$ $<$ 1.5 are somewhat higher than, but statistically consistent with, vdB+13 and Balogh et al.~(2016).  In Table 1 we summarise the above results as well as showing photometric cluster and field galaxy selection ranges and their corresponding co-moving volumes.

Our uncertainty calculations are performed in a more subtle way than in N+16a, combining random background and photometry samples into a single uncertainty rather than two added in quadrature, resulting in smaller error bars ultimately coming from the same two sources (background sampling and photometry).  The lower passive fraction and environmental quenching efficiency in our high-redshift bin than in N+16a can be attributed to the inclusion of only the three $z$ = 1.6 clusters in the highest redshift bin, whereas N+16a also included the z = 1.37 cluster.

We performed several checks on the robustness of our results, including exclusion of galaxies within 0.1 mag of the UVJ passive/star-forming line; using a 1 Mpc radius for all clusters; averaging and randomly bootstrap resampling environmental quenching efficiencies for individual clusters instead of the combined galaxy population of all clusters in a given bin; and using only galaxies more massive than 10$^{10.53}$ M$_{\sun}$.  All yielded a similar difference between the $z$ = 1.6 sample and the others, although some produced larger error bars.  The consistency of our qualitative results with these tests suggests that our $z$ $\sim$ 1.6 clusters do have a lower mean environmental quenching efficiency.

Darvish et al.~(2016) also find evolution in the passive fraction and quenching efficiency between $z$ $\sim$ 1 and $z$ $\sim$ 1.5 down to 10$^{10}$ M$_{\sun}$, but only at a 2 $\sigma$ level as opposed to our 3$\sigma$ confidence in quenching efficiency and $>$3$\sigma$ confidence in passive fraction (see Table 1).  Our SpARCS cluster analysis is distinct from Darvish et al.~(2016) in that our clusters are spectroscopically confirmed, and also drawn from a much larger survey: SpARCS covers about 26 times the co-moving volume of UltraVISTA/COSMOS (see Table 1), from which Darvish et al.~draw their results.  Our SpARCS cluster sample is therefore likely to include environments denser and rarer than the highest-density environments appearing in Darvish et al.~(2016), making our study complementary to theirs, extending the analysis to the extremes of environmental density.

\begin{table*}
\caption{Passive fractions, quenching efficiencies, and comoving volume by redshift}
\label{tab1}
\centering
\begin{tabular}{lccccccccc}
\hline\hline
Redshift & $z_{phot,min}$ & $z_{phot,max}$ & No. Clusters & $f_{p}$ & Quenching Eff. & $f_{p,field}$ & $\sigma_{field}$ & Com. Vol. Field & Com. Vol. SpARCS  \\ 
 & & & & & & & & 10$^6$ Mpc$^3$ & 10$^6$ Mpc$^3$ \\ 
\hline
z $\sim$ 0.95 & 0.77 & 1.14 & 6 & $0.88_{-0.03}^{+0.04}$ & $0.73_{-0.07}^{+0.08}$ & 0.552 & 0.025 & 3.21 & 83 \\
\\
z $\sim$ 1.3 & 1.04 & 1.48 & 5 & $0.80_{-0.09}^{+0.12}$ & $0.62_{-0.15}^{+0.21}$ & 0.473 & 0.025 & 4.69 & 121 \\
\\
z $\sim$ 1.6 & 1.47 & 1.76 & 3 & $0.42_{-0.13}^{+0.10}$ & $0.16_{-0.19}^{+0.15}$ & 0.317 & 0.032 & 4.37 & 113 \\

\hline  
\end{tabular}
\end{table*}

\section{Discussion}

The timespan between $z$ $=$ 1.6 and $z$ $\sim$ 1.3 is only $\sim$ 1 Gyr, indicating possibly rapid evolution in environmental quenching efficiency in this redshift range.  A dramatic change in quenching efficiency in a short time would make sense if, as Muldrew, Hatch, and Cooke (2015) predict, galaxy clusters in this redshift range collapse rapidly, at which point the intracluster medium would then be capable of stripping the galaxies' gas supplies.  However, halo age may also matter.  In Noble et al.~(2013) and Muzzin et al.~(2014), for instance, galaxies with dynamical evidence for having spent less time in the cluster were more likely to be star-forming or only very recently quenched.

The variety of quenched fractions in massive high-redshift clusters in the literature suggests that both halo mass and halo age may be important.  For instance, the Cooke et al.~(2016) cluster at $z$ $\sim$ 1.6 has a red fraction of massive galaxies comparable to $z$ $\sim$ 1 clusters, while the similarly massive $z$ $\sim$ 2.5 protocluster of Wang et al.~(2016) is still dominated by massive star-forming galaxies like its less-massive $z$ $>$ 2 counterparts.  The Quadri et al.~(2012) cluster at $z$ $\sim$ 1.6 has an intermediate passive fraction between these two extremes.  

In our data, considering only galaxies above the shallowest common stellar mass limit for individual clusters ($10^{10.53}$ M$_{\sun}$, 0.23 dex higher than our working limit of $10^{10.3}$ M$_{\sun}$), the difference between the largest and smallest environmental quenching efficiency is higher at $z$ = 1.6 than any other redshift.  This suggests that stochastic cluster-to-cluster variation in halo masses and ages may be especially important at the highest redshifts in determining environmental effects on cluster galaxies.

\section{Conclusions}

We conclude that for cluster galaxies more massive than 10$^{10.3}$ M$_{\sun}$, the evolution in the passive fraction in clusters outpaces that of the field, rising from $42_{-13}^{+10}$\% at $z$ $\sim$ 1.6 to $80_{-9}^{+12}$\% at $z$ $\sim$ 1.3 and $88_{-3}^{+4}$\% at $z$ $<$ 1.1.  The environmental quenching efficiency evolves substantially, jumping from $16_{-19}^{+15}$\% at $z$ $\sim$ 1.6 to $62_{-15}^{+21}$\% at $z$ $\sim$ 1.3 and $73_{-7}^{+8}$\% at $z$ $\lesssim$ 1.1.  Since our study selects galaxy clusters from similar comoving volumes using similar methods, and analyses each sub-sample in a nearly identical manner, we have high confidence that our results represent genuine evolution of galaxy clusters as they grow over cosmic time.

Our work is the first to show directly that environmental quenching becomes increasingly dominant between $z$ $=$ 1.6 and $z$ $=$ 0.9. In simulations of structure formation, this is an epoch in which the average cluster-mass halo grows especially rapidly (Muldrew, Hatch, and Cooke 2015) and is therefore very likely to become virialised in its central regions, suggesting a prerequisite for environmental quenching may be that galaxies are sub-halos in a virialised halo.  By extrapolation, our work also strongly indicates that quenching at $z$ $>$ 2 may be caused almost entirely by internal processes.

Our study leaves open the question of the relative extent to which cluster mass vs.~cluster age is responsible for environmental quenching.  Based on the variety of environmental quenching efficiencies and passive fractions reported in the $z$ $>$ 1.5 cluster literature and the variation within the SpARCS sample, we suggest that both halo mass and halo age are likely to play a role in the observed environmental quenching efficiency.

\section*{Acknowledgements}

This research uses observations obtained with MegaPrime/MegaCam, a joint project of CFHT and CEA/DAPNIA, at the Canada-France-Hawaii Telescope (CFHT) which is operated by the National Research Council (NRC) of Canada, the Institut National des Sciences de l'Univers of the Centre National de la Recherche Scientifique (CNRS) of France, and the University of Hawaii. This work is based in part on data products produced at TERAPIX and the Canadian Astronomy Data Centre as part of the Canada-France-Hawaii Telescope Legacy Survey, a collaborative project of NRC and CNRS. This project also uses observations taken at the ESO Paranal Observatory (ESO programs 085.A-0166, 085.A-0613, 086.A-0398, 087.A-0145, 087.A-0483, 088.A-0639, 089.A-0125, and 091.A-0478), Las Campanas Observatory in Chile, the Keck Telescopes in Hawaii, and the Spitzer Space Telescope, which is operated by the Jet Propulsion Laboratory, California Institute of Technology under a contract with NASA.

Our research team recognizes support from the following grants: Universidad Andres Bello internal research project DI-651-15/R (JN), BASAL Center for Astrophysics and Associated Technologies (CATA) and FONDECYT grant no. 1130528 (RD), European Research Council FP7 grant number 340519 (RFJvdB), NSF grant AST-1518257 (MCC), NSF grant AST-1517863, NASA programs GO-13306, GO-13677, GO-13747 \& GO-13845/14327 (GW).  The GO grants come from from the Space Telescope Science Institute, which is operated by AURA, Inc., under NASA contract NAS 5-26555.









\bsp	
\label{lastpage}
\end{document}